\documentclass[twocolumn,showpacs,floats,floatfix,superscriptaddress,aps,pra]{revtex4}
\usepackage{amsfonts}
\usepackage{amssymb}
\usepackage{amsmath}
\usepackage{graphicx}
\usepackage{bm}
\usepackage{color}
\usepackage{epsfig}
\usepackage{calc}
\usepackage{ifthen}

\begin{document}

\title{Broadband polarization retarder}
\author{A. A. Rangelov}
\affiliation{Department of Physics, Sofia University, James Bourchier 5 blvd., 1164
Sofia, Bulgaria}
\date{\today }

\begin{abstract}
We derive a new kind of effective broadband polarization device using the
analogy between the equation for the polarization of light, propagating
through an optically anisotropic medium, and the optical Bloch equation in
adiabatic regime, which describes coherent laser excitation of a two-state
atom.
\end{abstract}

\maketitle


Two simple and very convenient tools to describe the change of the
polarization of light, transmitted through anisotropic optical media, are
the Jones vector and Stokes vector \cite{Wolf,Azzam,Goldstein}. The equation
of motion for the Jones vector in a medium with zero polarization-dependent
loss has a Schr\"{o}dinger equation form \cite{Aben,Kubo78}, while the
equation of motion for the Stokes vector has a Bloch equation form \cite%
{Kubo80,Kubo81,Kubo83}. These properties have been used recently to draw
analogies between the motion of the Jones vector, alternatively the Stokes
vector, and a spin-1/2 particle in nuclear magnetic resonance, or
alternatively an optically driven two-state atom in quantum optics,
respectively described by Schr\"{o}dinger or Bloch equations \cite%
{Kuratsuji98,Zapasskii,Kuratsuji07,Rangelov,Botet08,Botet10}.

The phase delay between the fast and the slow eigenpolarizations of
traditional birefringent polarization devices scales in proportion
to the frequency of the light and thus such devices are frequency
dependent: a half wave plate and a quarter wave plate work for a
single frequency. At the same time, there exists a broadband
polarization transforming device, known as Fresnel rhomb
\cite{Wolf,Azzam,Goldstein}, but the quality of polarization
transformation is limited by the dispersive properties of the
material, the rhomb is made of. Except the Fresnel rhomb, broadband
polarization converters have proven to be inefficient, whereas
efficient polarization converters are narrowband.

The adiabatic polarization conversion proposed here is frequency
independent: any input polarization state is transformed to the
desired output polarization state regardless of the wavelength. This
conversion acts as a broadband device, which is limited only by the
absorptive characteristics of the device instead of its
birefringence bandwidth. It is noteworthy that the proposed
technique is analogous to the adiabatic technique in quantum optics
\cite{Allen,Sho90,Vit01a,Vit01b,Sho08}, and hence enjoys the same
advantages in terms of efficiency and robustness.

We begin with the derivation of the basic equations of light
polarization evolution in anisotropic nonmagnetic linear crystal
with negligible absorption. For such media, where no free charges
and no currents are present, the Maxwell equations are
\begin{subequations}
\label{Maxwell equations}
\begin{gather}
\nabla \cdot \left( \mathbf{\epsilon }\cdot \mathbf{E}\right) =0\,,
\label{Gauss's law} \\
\nabla \cdot \mathbf{B}=0,  \label{Gauss's law for magnetism} \\
\nabla \times \mathbf{E}=-\frac{\partial \mathbf{B}}{\partial t},
\label{Faraday's law} \\
\nabla \times \mathbf{B}=\frac{\mathbf{\epsilon }}{c^{2}}\cdot \frac{%
\partial \mathbf{E}}{\partial t},  \label{Ampere's law}
\end{gather}%
where $\mathbf{E}$ is the electric vector, $\mathbf{B}$ is the
magnetic vector, $\mathbf{\epsilon }$ is the electric permittivity
tensor and $c$ is the light velocity in vacuum. Writing the Maxwell
equations (\ref{Maxwell equations}) in a Cartesian coordinate system
for which $z$ direction, the direction of the propagation of light,
coincides with one of the optical axes of the  crystal, then the
tensor $\mathbf{\epsilon }$ can be written as
\cite{Botet08,Botet10}
\end{subequations}
\begin{equation}
\mathbf{\epsilon }\left( z\right) =\left[
\begin{array}{ccc}
\epsilon _{xx}\left( z\right) & \epsilon _{xy}\left( z\right) & 0 \\
\epsilon _{xy}\left( z\right) & \epsilon _{yy}\left( z\right) & 0 \\
0 & 0 & \epsilon _{z}\left( z\right)%
\end{array}%
\right] ,  \label{electric permittivity tensor}
\end{equation}%
where we assume that it is a symmetric function of $z$ only. The components $%
\epsilon _{xx}\left( z\right) ,\epsilon _{yy}\left( z\right) $ and $\epsilon
_{xy}\left( z\right) $ are expressed in terms of the principal values $%
\epsilon _{x}\left( z\right) $ and $\epsilon _{y}\left( z\right) $:
\begin{subequations}
\label{principal values}
\begin{eqnarray}
\epsilon _{xx}\left( z\right) &=&\frac{\epsilon _{x}\left( z\right)
-\epsilon _{y}\left( z\right) }{2}\cos \left( 2\varphi \right) \,, \\
\epsilon _{yy}\left( z\right) &=&\frac{\epsilon _{y}\left( z\right)
-\epsilon _{x}\left( z\right) }{2}\cos \left( 2\varphi \right) , \\
\epsilon _{xy}\left( z\right) &=&\frac{\epsilon _{y}\left( z\right)
-\epsilon _{x}\left( z\right) }{2}\sin \left( 2\varphi \right) ,
\end{eqnarray}%
where $\varphi $ is an angle of rotation between the xy plane of the chosen
Cartesian coordinate system and the optical axes.

If we take the curl of Eq. (\ref{Faraday's law}), using the vector identity $%
\nabla \times \left( \nabla \times \mathbf{E}\right) =\nabla \left( \nabla
\cdot \mathbf{E}\right) \mathbf{-}\Delta \mathbf{E}$, and if we assume that
the electric field varies slowly in the plane transverse to the propagation
direction, $\nabla \cdot \mathbf{E}\approx 0$, we arrive at the wave
equation for the electric field:
\end{subequations}
\begin{equation}
\nabla ^{2}\mathbf{E=}\frac{\mathbf{\epsilon }}{c^{2}}\cdot \frac{\partial
^{2}\mathbf{E}}{\partial t^{2}}.  \label{wave equation}
\end{equation}%
We shall deal with light propagation in one dimension. Let a monochromatic
plane wave with frequency $\omega $ propagate through the inhomogeneous
crystal along the $z$ axis:
\begin{equation}
\mathbf{E}\left( z,t\right) \mathbf{=}\left[
\begin{array}{c}
A_{x}\left( z\right) \mathbf{e}^{\left( ikz-\omega t\right) } \\
A_{y}\left( z\right) \mathbf{e}^{\left( ikz-\omega t\right) } \\
0%
\end{array}%
\right] .
\end{equation}%
Since the electric field $\mathbf{E}\left( z,t\right) $ depends only on the
longitudinal coordinate $z$, we can replace $\nabla ^{2}$\ by $\partial
^{2}/\partial z^{2}$, and as a result we obtain the following simplified
form of Eq. (\ref{wave equation}), written as two scalar equations:
\begin{subequations}
\label{two scalar equations}
\begin{eqnarray}
k^{2}A_{x}-\frac{\partial ^{2}A_{x}}{\partial z^{2}}-2ik\frac{\partial A_{x}%
}{\partial z} &=&\frac{\omega ^{2}}{c^{2}}\left( \epsilon
_{xx}A_{x}+\epsilon _{xy}A_{y}\right) , \\
k^{2}A_{y}-\frac{\partial ^{2}A_{y}}{\partial z^{2}}-2ik\frac{\partial A_{y}%
}{\partial z} &=&\frac{\omega ^{2}}{c^{2}}\left( \epsilon
_{xy}A_{x}+\epsilon _{yy}A_{y}\right) .
\end{eqnarray}%
It is generally acceptable to neglect the second term on the
left-hand side of each equation, as this term is much smaller
compared to the third one. This is known as the slowly varying
amplitude approximation \cite{Boyd} and is valid whenever we have
\end{subequations}
\begin{equation}
\left\vert \frac{\partial ^{2}A_{x,y}}{\partial z^{2}}\right\vert \ll
\left\vert 2k\frac{\partial A_{x,y}}{\partial z}\right\vert .
\end{equation}%
This condition requires the fractional change at a distance of the order of
an optical wavelength to be much smaller than unity. In this case Eq. (\ref%
{two scalar equations}) becomes
\begin{equation}
i\frac{\partial }{\partial z}\left[
\begin{array}{c}
A_{x} \\
A_{y}%
\end{array}%
\right] =\frac{\omega }{2c\sqrt{\epsilon _{z}}}\left[
\begin{array}{cc}
\epsilon _{z}-\epsilon _{xx} & -\epsilon _{xy} \\
-\epsilon _{xy} & \epsilon _{z}-\epsilon _{yy}%
\end{array}%
\right] \left[
\begin{array}{c}
A_{x} \\
A_{y}%
\end{array}%
\right] ,
\end{equation}%
where we have taken into account that $k^{2}=\left( \omega ^{2}\epsilon
_{z}\right) /c^{2}$. We note that the repeating diagonal term $\epsilon _{z}$
can be removed from the last equation by incorporating it as an identical
phase in the amplitudes $A_{x}$ and $A_{y}$.

Considering Eq. (\ref{principal values}), we rewrite the last equation as
\begin{equation}
i\frac{\partial }{\partial z}\left[
\begin{array}{c}
A_{x} \\
A_{y}%
\end{array}%
\right] =\frac{1}{2}\left[
\begin{array}{cc}
-\Delta & \Omega \\
\Omega & \Delta%
\end{array}%
\right] \left[
\begin{array}{c}
A_{x} \\
A_{y}%
\end{array}%
\right] ,  \label{two-state atom}
\end{equation}%
where
\begin{subequations}
\label{coupling and detuning}
\begin{eqnarray}
\Omega &=&\mu \sin \left( 2\varphi \right) ,  \label{Rabi frequency} \\
\Delta &=&\mu \cos \left( 2\varphi \right) ,  \label{Detuning} \\
\mu &=&\frac{\omega \left( \epsilon _{x}-\epsilon _{y}\right) }{2c\sqrt{%
\epsilon _{z}}}.
\end{eqnarray}%
If we map coordinate dependance into time dependance, Eq. (\ref{two-state
atom}) is equivalent to the Schr\"{o}dinger equation for a two-state atom in
rotating-wave approximation \cite{Allen,Sho90,Vit01a,Vit01b,Sho08}, where $%
A_{x}$ and $A_{y}$\ are the probability amplitudes for the ground state
(horizontal polarization) and the excited state (vertical polarization). The
off-diagonal element $\Omega $ in Eq. (\ref{two-state atom}) is known as
Rabi frequency, while the element $\Delta $ corresponds to the atom-laser
detuning \cite{Allen,Sho90}.

We can express the (real-valued) components of the Stokes vector \cite%
{Wolf,Azzam,Goldstein} by the components of Jones vector:
\end{subequations}
\begin{subequations}
\begin{eqnarray}
S_{0} &=&\left\vert A_{x}\right\vert ^{2}+\left\vert A_{y}\right\vert ^{2},
\\
S_{1} &=&\left\vert A_{x}\right\vert ^{2}-\left\vert A_{y}\right\vert ^{2},
\\
S_{2} &=&A_{y}A_{x}^{\ast }+A_{x}A_{y}^{\ast }, \\
S_{3} &=&i\left( A_{x}A_{y}^{\ast }-A_{y}A_{x}^{\ast }\right) .
\end{eqnarray}%
Since the medium has negligible absorption, the value of $S_{0}$ is
conserved and therefore Eq. (\ref{two-state atom}) turns into the optical
Bloch equation \cite{Allen,Sho90}:
\end{subequations}
\begin{equation}
\frac{\partial }{\partial z}\left[
\begin{array}{c}
S_{1} \\
S_{2} \\
S_{3}%
\end{array}%
\right] =\left[
\begin{array}{ccc}
0 & 0 & \Omega \\
0 & 0 & \Delta \\
-\Omega & -\Delta & 0%
\end{array}%
\right] \left[
\begin{array}{c}
S_{1} \\
S_{2} \\
S_{3}%
\end{array}%
\right] .  \label{Stokes}
\end{equation}

Traditional polarization devices work on resonance ($\Delta =0$), which is
achieved by adjusting the plane of the incident light, so that it makes an
angle $\varphi =\pi /4$ with the fast axis (see Eq. (\ref{Detuning})).
Furthermore, to achieve good polarization conversion an exact area condition
is required ($\int \Omega \,dz=n\pi $ with $n$ being half integer).
Consequently, the traditional retarders depend on the propagation length
(the thickness of the plate) and are not broadband. Hence they are not
optimal. To this end, we propose an alternative and robust adiabatic
technique to create broadband polarization retarder.

Let us assume that we use layers of the same material to make our retarder
and that we change only the angle $\varphi $ from layer to layer. Then we
can write Eq. (\ref{Stokes}), using Eq. (\ref{coupling and detuning}), in
the so-called adiabatic basis \cite{Allen,Sho90,Vit01a,Vit01b,Sho08}:
\begin{equation}
\frac{\partial }{\partial z}\left[
\begin{array}{c}
S_{1}^{A} \\
S_{2}^{A} \\
S_{3}^{A}%
\end{array}%
\right] =\left[
\begin{array}{ccc}
i\mu & \frac{2}{\sqrt{2}}\frac{\partial \varphi }{\partial z} & 0 \\
-\frac{2}{\sqrt{2}}\frac{\partial \varphi }{\partial z} & 0 & -\frac{2}{%
\sqrt{2}}\frac{\partial \varphi }{\partial z} \\
0 & \frac{2}{\sqrt{2}}\frac{\partial \varphi }{\partial z} & -i\mu%
\end{array}%
\right] \left[
\begin{array}{c}
S_{1}^{A} \\
S_{2}^{A} \\
S_{3}^{A}%
\end{array}%
\right] .  \label{adiabatic Stokes}
\end{equation}%
The connection between the Stokes vector $\mathbf{S}\left( z\right) $ in the
original basis and the Stokes vector $\mathbf{S}^{A}\left( z\right) $ in the
adiabatic basis is given by
\begin{equation}
\mathbf{S}\left( z\right) =R\left( z\right) \mathbf{S}^{A}\left( z\right) ,
\end{equation}%
with $R\left( z\right) $ being the unitary transformation matrix
\begin{equation}
R\left( z\right) =\frac{1}{\sqrt{2}}\left[
\begin{array}{ccc}
\sin \left( 2\varphi \right) & \sqrt{2}\cos \left( 2\varphi \right) & \sin
\left( 2\varphi \right) \\
\cos \left( 2\varphi \right) & -\sqrt{2}\sin \left( 2\varphi \right) & \cos
\left( 2\varphi \right) \\
i & 0 & -i%
\end{array}%
\right] .
\end{equation}

For adiabatic evolution of the system there are no transitions between the
amplitudes $S_{1}^{A}$, $S_{2}^{A}$ and $S_{3}^{A}$. Hence $\left\vert
S_{1}^{A}\right\vert ,\left\vert S_{2}^{A}\right\vert $ and $\left\vert
S_{3}^{A}\right\vert $ remain constant \cite{Allen,Sho90,Vit01a,Vit01b,Sho08}%
. Mathematically, adiabatic evolution means that in Eq. (\ref{adiabatic
Stokes}) the non-diagonal terms can be neglected compared to the diagonal
terms, which holds when we have \cite{Allen,Sho90,Vit01a,Vit01b,Sho08}
\begin{equation}
\left\vert \frac{\partial \varphi }{\partial z}\right\vert \ll \left\vert
\mu \right\vert .
\end{equation}%
Thus, adiabatic evolution requires smooth $z$-dependance of the angle $%
\varphi $ and large rotary power $\left\vert \mu \right\vert $. For a pure
adiabatic evolution, the solution of Eq. (\ref{adiabatic Stokes}) is very
simple,%
\begin{equation}
\mathbf{S}^{A}\left( z_{f}\right) =U^{A}\left( z_{f},z_{i}\right) \mathbf{S}%
^{A}\left( z_{i}\right) ,
\end{equation}%
where the adiabatic propagator $U^{A}\left( z_{f},z_{i}\right) $ is diagonal
and contains only phase factors:%
\begin{equation}
U^{A}\left( z_{f},z_{i}\right) =\left[
\begin{array}{ccc}
\exp \left( i\eta \right) & 0 & 0 \\
0 & 1 & 0 \\
0 & 0 & \exp \left( -i\eta \right)%
\end{array}%
\right] ,
\end{equation}%
with the adiabatic phase $\eta =\int_{z_{i}}^{z_{f}}\mu dz$. The propagator
in the original basis, the reduced Mueller matrix, is%
\begin{equation}
U\left( z_{f},z_{i}\right) =R\left( z_{f}\right) U^{A}\left(
z_{f},z_{i}\right) R^{\dagger }\left( z_{i}\right) ,
\end{equation}%
or explicitly:
\begin{widetext}
\begin{equation}
U\left( z_{f},z_{i}\right) =\left[
\begin{array}{ccc}
\cos \left( \alpha \right) \cos \left( \beta \right) +\sin \left( \alpha
\right) \sin \left( \beta \right) \cos \left( \eta \right)  & -\sin \left(
\alpha \right) \cos \left( \beta \right) +\cos \left( \alpha \right) \sin
\left( \beta \right) \cos \left( \eta \right)  & \sin \left( \beta \right)
\sin \left( \eta \right)  \\
\cos \left( \beta \right) \sin \left( \alpha \right) \cos \left( \eta
\right) -\cos \left( \alpha \right) \sin \left( \beta \right)  & \cos \left(
\alpha \right) \cos \left( \beta \right) \cos \left( \eta \right) +\sin
\left( \alpha \right) \sin \left( \beta \right)  & \cos \left( \beta \right)
\sin \left( \eta \right)  \\
-\sin \left( \alpha \right) \sin \left( \eta \right)  & -\cos \left( \alpha
\right) \sin \left( \eta \right)  & \cos \left( \eta \right)
\end{array}%
\right], \label{adiabatic solution}
\end{equation}
with $\alpha =2\varphi \left( z_{i}\right) $ and $\beta =2\varphi
\left( z_{f}\right) $.
\end{widetext}

Equation (\ref{adiabatic solution}) gives the general adiabatic
evolution scenario for an arbitrary polarization. Naturally and
without loss of generality we can fix $x$ and $y$ axis of the
Cartesian coordinate system to coincide with the fast and slow
optical axis for the first layer of the
crystal, which is equivalently to set $\varphi (z_{i})=0$. In this case Eq. (%
\ref{adiabatic solution}) reduces to simpler form:%
\begin{equation}
U\left( z_{f},z_{i}\right) =\left[
\begin{array}{ccc}
\cos \left( \beta \right) & \sin \left( \beta \right) \cos \left( \eta
\right) & \sin \left( \beta \right) \sin \left( \eta \right) \\
-\sin \left( \beta \right) & \cos \left( \beta \right) \cos \left( \eta
\right) & \cos \left( \beta \right) \sin \left( \eta \right) \\
0 & -\sin \left( \eta \right) & \cos \left( \eta \right)%
\end{array}%
\right] .  \label{Mueller matrix}
\end{equation}%
Obviously there are two terms in the last equation that do not depend from
the adiabatic phase $\eta $, therefore those terms are not frequency
dependent and a retarder, that use only those two terms, will be frequency
independent.

\emph{\textbf{\ Broadband rotation for linearly polarized light:}} If
initially the light is linearly polarized in horizontal direction, $\mathbf{S%
}(z_{i})=(1,0,0)$, then from Eq. (\ref{Mueller matrix}) we have
\begin{eqnarray}
S_{1}\left( z_{f}\right) &=&\cos \left( 2\varphi \left( z_{f}\right) \right)
, \\
S_{2}\left( z_{f}\right) &=&-\sin \left( 2\varphi \left( z_{f}\right)
\right) , \\
S_{3}\left( z_{f}\right) &=&0.
\end{eqnarray}%
Therefore we can end up with a vertically polarized light, $\mathbf{S}%
(z_{f})=(-1,0,0)$, if we set the final angle $\varphi (z_{f})=\pi /2$. This
process is reversible: if we start with a vertically polarized light and set
the final angle to be $\varphi (z_{f})=\pi /2$, we achieve reversal of the
direction of motion and we end up with a horizontal polarization.

Analogously, if initially the light is linearly polarized in horizontal
direction, $\mathbf{S}(z_{i})=(1,0,0)$, but now we choose the final angle to
be $\varphi (z_{f})=3\pi /4$ or $\varphi (z_{f})=\pi /4$, then we end up
either with a linear $+45^{\circ }$ polarized light, $\mathbf{S}%
(z_{f})=(0,1,0)$, or with a linear $-45^{\circ }$ polarized light, $\mathbf{S%
}(z_{f})=(0,-1,0)$. Again, those processes are reversible.

The presented adiabatic retarder is advantageous because the polarization
rotation depends only on the angle of rotation. These retarder is frequency
independent and it is robust against variations of the propagation length,
rotary power, etc., in contrast to the traditional retarders.

In conclusion, we have shown that using the analogy between the equation,
which describes the polarization state of light, propagating through an
optically anisotropic medium, and the optical Bloch equation in adiabatic
regime, which describes coherent laser excitation of a two-state atom, it is
possible to build effective and broadband polarization retarder.

This work has been supported by the European Commission project
FASTQUAST, the Bulgarian NSF grants D002-90/08, DMU02-19/09 and
Sofia University Grant 022/2011. The author is grateful to N. V.
Vitanov for stimulating discussions.

\end{document}